\documentclass[a4paper,11pt]{article}
\usepackage{amsmath,amsthm,amssymb,mathrsfs,graphicx,braket,empheq,color}
\usepackage[titletoc,title]{appendix}
\usepackage{comment} 

\newcommand{\nc}{\newcommand}
\nc{\rnc}{\renewcommand}
\nc{\nn}{\nonumber}
\nc{\der}{{\partial}}
\rnc{\Im}{{\textrm{Im}\,}}
\rnc{\Re}{{\textrm{Re}\,}}
\nc{\db}{\displaybreak[0]\\}
\nc{\astl}{\underset{\mathcal{L}_l}{\ast}}
\nc{\astj}{\underset{\mathcal{L}_j}{\ast}}
\nc{\astll}{\underset{\mathcal{L}_1}{\ast}}
\nc{\Yt}{Y^{\textrm{th}}}
\nc{\rmd}{{\rm d}}
\nc{\rmi}{{\rm i}}
\nc{\rme}{{\rm e}}
\nc{\tcr}{\textcolor{red}}

\nc{\End}{\mathrm{End}}

\theoremstyle{definition}

\numberwithin{equation}{section}

\begin{document}

\title{Quantum circuit for the fast Fourier transform}

\author{
Ryo Asaka\thanks{E-mail: 1219502@ed.tus.ac.jp}, \,
Kazumitsu Sakai\thanks{E-mail: k.sakai@rs.tus.ac.jp} \, and
Ryoko Yahagi \thanks{E-mail: yahagi@rs.tus.ac.jp}
\\\\
\textit{Department of Physics,
Tokyo University of Science,}\\
 \textit{Kagurazaka 1-3, Shinjuku-ku, Tokyo, 162-8601, Japan} \\
\\\\
\\
}

\date{November 8, 2019}

\maketitle

\begin{abstract}
We propose an implementation of the algorithm for the fast Fourier 
transform (FFT) as a quantum circuit consisting of a combination of 
some quantum gates. In our implementation, a data sequence is expressed 
by a tensor product of vector spaces. 
Namely, our FFT is defined as a transformation of the tensor product of
quantum states. It is essentially different from the so-called quantum 
Fourier transform (QFT) defined to be a linear transformation 
of the amplitudes for the superposition of quantum states. The quantum circuit for the 
FFT consists of several circuits for elementary arithmetic operations 
such as a quantum adder, subtractor and shift operations, which are 
implemented as effectively as possible. Namely, our circuit does not
generate any garbage bits. The advantages of our method compared to the QFT 
are its high versatility, and data storage efficiency in terms, for 
instance, of the quantum image processing.

\end{abstract}

\maketitle
\section{Introduction}

Quantum computing, which utilizes quantum entanglement and
quantum superpositions inherent to quantum mechanics, is 
rapidly gaining ground to overcome the limitations of classical
computing. Shor's algorithm \cite{Shor} solving the integer 
factorization problem in a polynomial time and Grover's algorithm 
\cite{Grover} making it possible to substantially speed up 
the search in unstructured databases\footnote{In fact, an 
effective encoding method to convert
classical data to quantum states (e.g. a quantum version
of random access memory (qRAM)) \cite{qRAM1,qRAM2,qRAM3} 
is necessary to take advantage of quantum computing.}
 are one of the best-known 
examples of the astounding properties of quantum computing
(see \cite{QCQI}, for example, for various applications of 
quantum computing).

An implementation of the Fourier transform as a quantum circuit
sometimes plays a crucial role on quantum computing.
Indeed, the quantum Fourier transform (QFT) \cite{QFT} is a key  
ingredient of many important quantum algorithms, including
Shor's  factoring algorithm and the quantum phase estimation algorithm
to estimate the eigenvalues of a unitary operator.
Here, the QFT is  
the Fourier transform for the amplitudes of a quantum state:
\begin{equation}
\sum_{j=0}^{N-1} x_j \ket{j} \longmapsto \sum_{k=0}^{N-1} X_k \ket{k},
\label{super-QFT}
\end{equation}
where we set $N=2^n$, and 
the amplitudes $\{X_k\}$ are the classical discrete Fourier transform 
of the amplitudes $\{x_j\}$
\begin{equation}
X_k=\sum_{j=0}^{N-1}W_N^{j k}x_j,
\qquad 
x_j=\frac{1}{N}\sum_{k=0}^{N-1} W_N^{-jk}X_k,
\label{FT}
\end{equation}
where $W_N:=\exp(-2\pi i/N)$.
Due to the superposition of the state \eqref{super-QFT} and quantum 
parallelism, the QFT can be implemented in a quantum circuit 
consisting of $O(n^2)$ quantum gates, which is much more efficient 
than the fast Fourier transform (FFT) \cite{FFT} whose complexity 
of the computation is $O(n 2^n)$.

The Fourier transform that we consider in this paper is somewhat 
different from the QFT: We propose a quantum implementation of the 
algorithm of the FFT rather than the QFT. In our procedure, a data sequence 
is expressed in terms of a tensor product of vector spaces:
$\bigotimes_{j=0}^{N-1}\ket{x_j}$. Namely, the state vectors representing 
the given classical information are prepared via so-called {\it basis} encoding 
\cite{Schuld}. (On the other hand, the QFT \eqref{super-QFT} is based on the 
{\it amplitude} encoding.) Based on the basis encoding, the Fourier transform 
is defined as
\begin{equation}
\bigotimes_{j=0}^{N-1}\ket{x_j}\longmapsto\bigotimes_{k=0}^{N-1}\ket{X_k},
\label{QFFT}
\end{equation}
where the data sequence  $\{X_k\}$ is the 
Fourier transform of $\{x_j\}$ as expressed in \eqref{FT}.
We adopt the reversible FFT \cite{FFT-Int} as an algorithm of 
the above Fourier transform
and implement it as a quantum circuit whose computational complexity
is $O(n 2^n)$. In this point of view, the processing speed is the same
as the classical one, as long as we consider only a single data
sequence. Nevertheless, there are following advantages compared 
to the classical FFT, and even compared to the QFT. 
The first is due to quantum parallelism. Namely, utilizing 
quantum superposition of multiple data sets, we can simultaneously 
process them. Note here that, there exist several problems  how to
encode classical data in quantum states (and also how to
read resultant superposed quantum data), which are peculiar
to quantum computing. To take advantage of quantum computing,
a qRAM suitable for quantum computation \cite{qRAM1,qRAM2,qRAM3} is necessary
(see Sect.~\ref{cost} for comparison
of computational costs between the 
classical FFT and our  quantum version of the FFT (let us denote it as QFFT)
including  data encoding).
The second is due to its high versatility:
The method is always applicable to data sets that can be processed 
by the conventional FFT. The third advantage is its data storage efficiency 
in terms, for instance, of the quantum image (see \cite{QFT-Amp1,QFT-Amp2,QFT-Amp3,
QFT-inBase} for some applications of the QFT to quantum data sets).

\begin{figure}[t]
\centering
\includegraphics[width=0.6\textwidth]{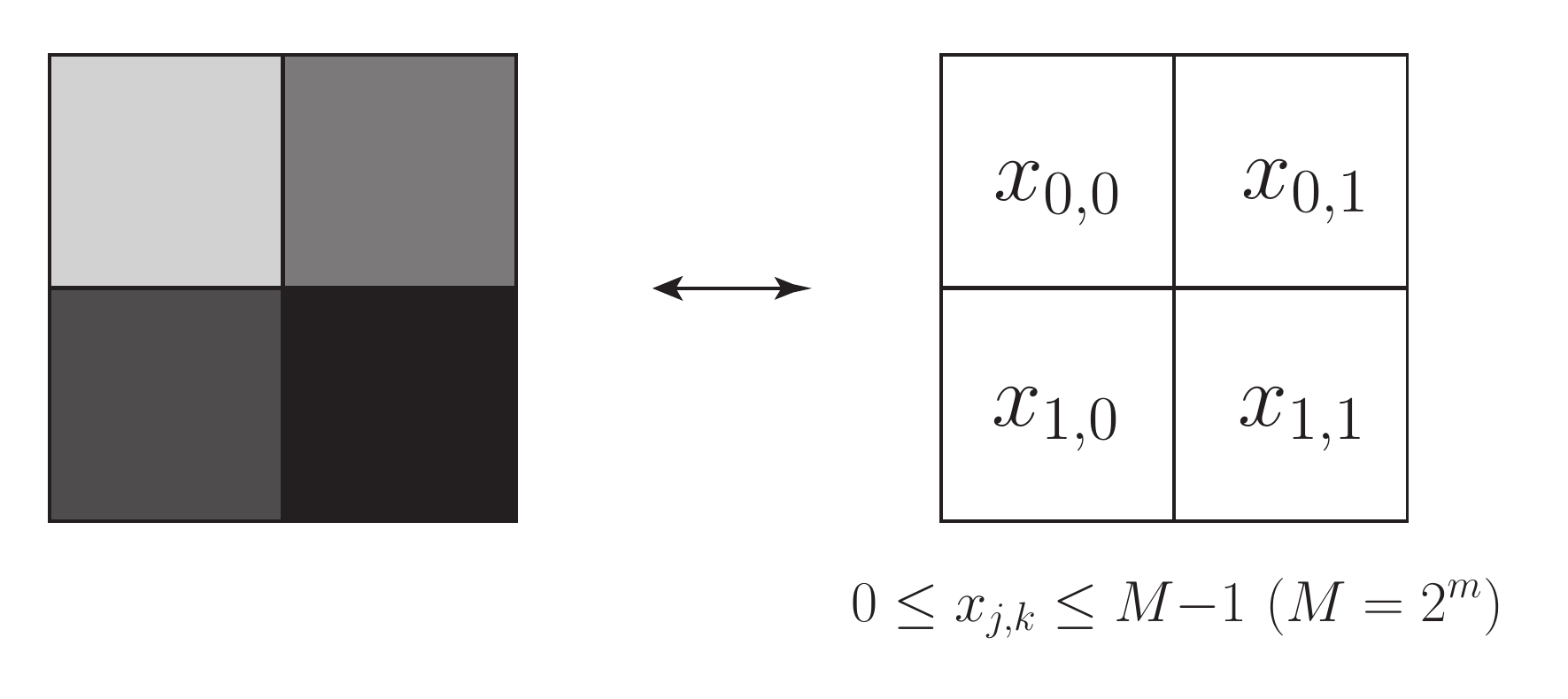}
\caption{
A $2\times 2$ pixel image with a grayscale value ranging from 0 to $M-1$ 
$(M=2^m)$ is depicted in the left panel. $x_{j,k}$ in the right panel denotes the 
value of the grayscale on the $(j,k)$-site. The quantum image can be
represented as 
$\ket{x_{0,0}}\otimes \ket{x_{0,1}}\otimes\ket{x_{1,0}}\otimes \ket{x_{1,1}}
\in (\mathbb{C}^2)^{\otimes 4m}$.
}
\label{gray}
\end{figure}

Let us illustrate the third advantage above with a simple example:
an $L \times L$ pixel image with a grayscale value ranging from 0 to $M-1$ 
($M=2^m$)
(see Fig. \ref{gray} for $L=2$).  (This problem is equivalent to a
lattice quantum many-body problem on an $L\times L$ square lattice with each 
site occupied by a particle with $M$ degrees of freedom.)
This quantum image $\ket{\psi^{(\alpha)}}$ ($\alpha$ denotes the label 
of the image) can be represented by a 
tensor product of vector spaces \cite{BaseRep1,BaseRep2,BaseRep3,BaseRep4}:
\begin{equation}
\ket{\psi^{(\alpha)}}=\bigotimes_{(j,k)=(0,0)}^{(L-1,L-1)} \ket{x^{(\alpha)}_{j,k}}, 
\qquad 0\le x^{(\alpha)}_{j,k}\le M-1.
\label{q-image}
\end{equation}
Since $\ket{\psi^{(\alpha)}}\in (\mathbb{C}^2)^{\otimes m L^2}$, 
it uses $mL^2$ qubits.
By use of the quantum superposition, the QFFT can simultaneously 
process at most $2^{mL^2}$ 
quantum images:
\begin{equation}
\ket{\Psi}=\sum_{\alpha=1}^{2^{mL^2}} c_{\alpha} \ket{\psi^{(\alpha)}}
\in (\mathbb{C}^2)^{\otimes m L^2}
 \qquad 
(c_{\alpha}\in \mathbb{C}).
\label{super-QFFT}
\end{equation}
On the other hand, 
to apply the QFT to the above image processing, we need to prepare the quantum image 
in the form of
\begin{equation}
\ket{\widetilde{\psi}^{(\alpha)}}=\sum_{(j,k)=(0,0)}^{(L-1,L-1)}
x^{(\alpha)}_{j,k}\ket{j,k},
\label{QFT-image}
\end{equation}
where $\ket{\widetilde{\psi}^{(\alpha)}}\in (\mathbb{C}^2)^{\otimes 2\log_2L}$ which 
uses only $2\log_2L$ qubits (cf. \eqref{q-image} for the QFFT) \cite{AmpRep1,AmpRep2,
AmpRep3}.  
However, since the Fourier coefficients (see \eqref{super-QFT} and \eqref{FT}) 
are expressed as the amplitudes of the
superposition, it takes exponentially long time  to extract all of them completely.
Furthermore, to properly perform the Fourier transform for the 
multiple $2^{m L^2}$ quantum images, they must be represented as
\begin{equation}\
\ket{\widetilde{\Psi}}=
\bigotimes_{\alpha=1}^{2^{mL^2}}\ket{\widetilde{\psi}^{(\alpha)}}
\in (\mathbb{C}^2)^{\otimes  (2^{mL^2+1})\log_2L}.
\label{dataset-QFT}
\end{equation}
Namely, for the QFT, $(2^{mL^2+1})\log_2L$ qubits are required to process the 
$2^{m L^2}$ quantum 
images, which are much larger than $m L^2$ qubits for the QFFT. 
Moreover, the QFT must be 
applied to  each  image individually, since the data set \eqref{dataset-QFT}
is not a superposition of images but a tensor product of each image\footnote{
A quantum superposition of images appropriate to the QFT can  be represented as 
$\bigoplus_{\alpha=1}^{2^{ m L^2}} \ket{\widetilde{\psi}^{(\alpha)}} \in 
(\mathbb{C}^2)^{2\log_2L+ m L^2}$
which uses $2\log_2L+ mL^2$ qubits. However, the resulting Fourier image
obtained by the QFT is no longer the same as any of the 
Fourier images of the individual quantum images.
}. As a result,  
the total processing time for the QFFT is 
shorter than that for the QFT, when the number of the quantum 
images is sufficiently large.

In this paper, we construct a quantum circuit of the above explained QFFT,
by implementing some elementary arithmetic operations such as 
a quantum adder \cite{Q-Adder1, Q-Adder2, Q-Adder3, Q-Adder4, Q-Adder5, Q-Adder6}, subtractor \cite{Q-Subb1, Q-Subb2, Q-Subb3, Q-Subb4} and  newly developed 
shift-type operations, as efficiently as possible: Our quantum 
circuit does not generate any garbage bits. 

The outline of the paper is as follows. 
In the subsequent section, introducing the algorithm of a quantum version of the 
FFT, we show the elementary arithmetic operations required for the implementation
of the QFFT as a quantum circuit. In Sect. 3, we actually implement these
elementary arithmetic operations into quantum circuits. In Sect. 4, 
combining these elementary circuits efficiently, 
we construct a quantum circuit for the QFFT.
The number of quantum gates required
for the implementation of the QFFT is estimated in Sect. 5.
The computational costs between the classical
FFT and the QFFT including data encoding are also discussed in this section.
In Sect. 6, we illustrate a concrete example of an application of 
the QFFT.
Sect. 7 is devoted to  a summary and discussion.
Some technical details are deferred to  Appendix.

%
\section{Elementary operations required for the QFFT}
\label{section:FFT}
In this section, we introduce the algorithm of a quantum version of the
FFT and pictorially represent several arithmetic operations
required for the implementation of the QFFT as a quantum circuit. 
(See \cite{FFT-app}, for instance, for the detailed algorithm of the FFT.)
We only use the basis encoding method to obtain the quantum states.
The matrix-like notations introduced here are helpful for the
implementation of quantum algorithms.

\subsection{Algorithm of the QFFT}
Let us start the formula \eqref{FT} and \eqref{QFFT} of the Fourier transform. 
Setting  $W_{N}=\exp(-2\pi i/N)$ $(N=2^n)$ and decomposing the summation 
in \eqref{FT} into the odd and even parts, we have
\begin{equation}
\Ket{X_k}=\Ket{G^{(n-1,0)}_k+W_{N}^{k} G^{(n-1,1)}_k}, 
\,\,
\Ket{X_{k+N/2}}=\Ket{G^{(n-1,0)}_k-W_{N}^k G^{(n-1,1)}_k},
\label{decompose}
\end{equation}
where $0\le k \le N/2-1$, and 
${G_k^{(n-1,p)}}$ ($p=0,1$) is the Fourier coefficients 
for $\{x_{2r+p}\}$  $(0\le r \le N/2-1)$:
\begin{equation}
{G^{(n-1,p)}_k}={\sum_{r=0}^{N/2-1}W_{N}^{2rk} x_{2r+p}} \quad (p=0,1).
\end{equation}
Note that $G^{(n-1,p)}_{k+N/2}=G^{(n-1,p)}_k$ 
and $W^{N/2}_N=-1$ hold.
In general, $X_k$ is a complex number and the notation $\Ket{X_k}$ stands for 
$\Ket{(X_k)_r}\otimes \Ket{(X_k)_i}$,
where $(X_k)_r$ and $(X_k)_i$ are the real and imaginary part of $X_k$, respectively.
Pictorially, \eqref{decompose} can be represented as
so-called a butterfly diagram:
\begin{equation}
\includegraphics[width=0.6\textwidth]{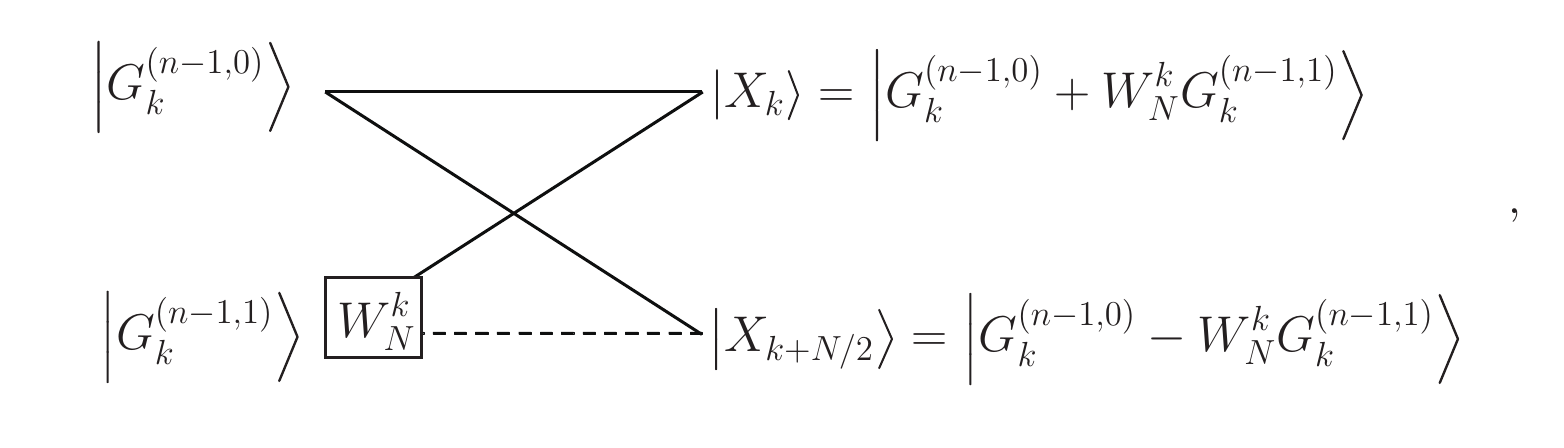}
\label{FT-pic}
\end{equation}
where $0\le k \le N/2-1$. Here, the broken line means
the multiplication  by $-1$.
For convenience, we also denote it as a matrix-like notation:
\begin{equation}
\begin{bmatrix}
\ket{X_k} \\[3mm]
\ket{X_{k+N/2}}
\end{bmatrix}
=\begin{bmatrix}
1 & 1 \\
1 &-1
\end{bmatrix}
\begin{bmatrix}
1 & 0 \\
0 &W_N^k
\end{bmatrix}
\begin{bmatrix}
\Ket{G^{(n-1,0)}_k} \\[3mm]
\Ket{G^{(n-1,1)}_k}
\end{bmatrix}
=\begin{bmatrix}
\Ket{G^{(n-1,0)}_k+W_{N}^{k} G^{(n-1,1)}_k}\\[3mm]
\Ket{G^{(n-1,0)}_k-W_{N}^{k} G^{(n-1,1)}_k}
\end{bmatrix}.
\label{matrix}
\end{equation}
Here, the matrix-like operation is  defined as
\begin{equation}
\begin{bmatrix}
A & B \\
C& D
\end{bmatrix}
\begin{bmatrix}
\ket{a} \\
\ket{b}
\end{bmatrix}
=\begin{bmatrix}
\ket{A a+B b}\\
\ket{C a+D d}
\end{bmatrix}
\quad
(A,B,C, D\in \mathbb{C}).
\label{matrix2}
\end{equation}
Do not confuse the above manipulation with conventional matrix operations:
The results are {\it not}  liner combinations of $\ket{a}$
and $\ket{b}$. The matrix-like notations are useful to implement
quantum algorithms as quantum circuits.

Again we decompose the Fourier transform for $\{x_{2r}\}$ (resp. $\{x_{2r+1}\}$)
into that for $\{x_{4s}\}$ and $\{x_{4s+2}\}$
(resp. $\{x_{4s+1}\}$ and $\{x_{4s+3}\}$) ($0\le s \le N/4-1$).
The result reads
\begin{equation}
\begin{bmatrix}
\Ket{G_k^{(n-1,p)}}\\[3mm]
\Ket{G_{k+N/4}^{(n-1,p)}}
\end{bmatrix}
=
\begin{bmatrix}
\Ket{G_k^{(n-2,p)}+W_{N/2}^k G_k^{(n-2,p+2)}}\\[3mm]
\Ket{G_k^{(n-2,p)}-W_{N/2}^k G_k^{(n-2,p+2)}}
\end{bmatrix}\quad
(p=0,1;0\le k \le N/4-1),
\end{equation}
where
\begin{equation}
G_k^{(n-2,q)}=\sum_{s=0}^{N/4-1} W_N^{4 sk} x_{4s+q} \quad (0\le q \le 3; \,\, 0\le s \le N/4-1).
\end{equation}
Repeating this procedure,  one obtains the following recursion relation:
\begin{equation}
\begin{bmatrix}
\Ket{G_k^{(n-m,p)}}\\[3mm]
\Ket{G_{k+N/2^{m+1}}^{(n-m,p)}}
\end{bmatrix}
=
\begin{bmatrix}
\Ket{G_k^{(n-m-1,p)}+W_{N/2^m}^k G_k^{(n-m-1,p+2^{m})}}\\[3mm]
\Ket{G_k^{(n-m-1,p)}-W_{N/2^m}^k G_k^{(n-m-1,p+2^{m})}}
\end{bmatrix},
\label{recursion}
\end{equation}
where $0\le p\le 2^{m}-1$, $0\le k \le N/2^{m+1}-1$. The
initial states are given by
\begin{equation}
\Ket{G_0^{(0,p)}}=\ket{x_p} \,\, (0\le p\le N-1).
\label{initial}
\end{equation}
This is the algorithm of the  QFFT. The classical version is
reproduced by just interpreting the state vectors as  scalars.

\begin{figure}[t]
  \centering
  \includegraphics[width=0.7\textwidth]{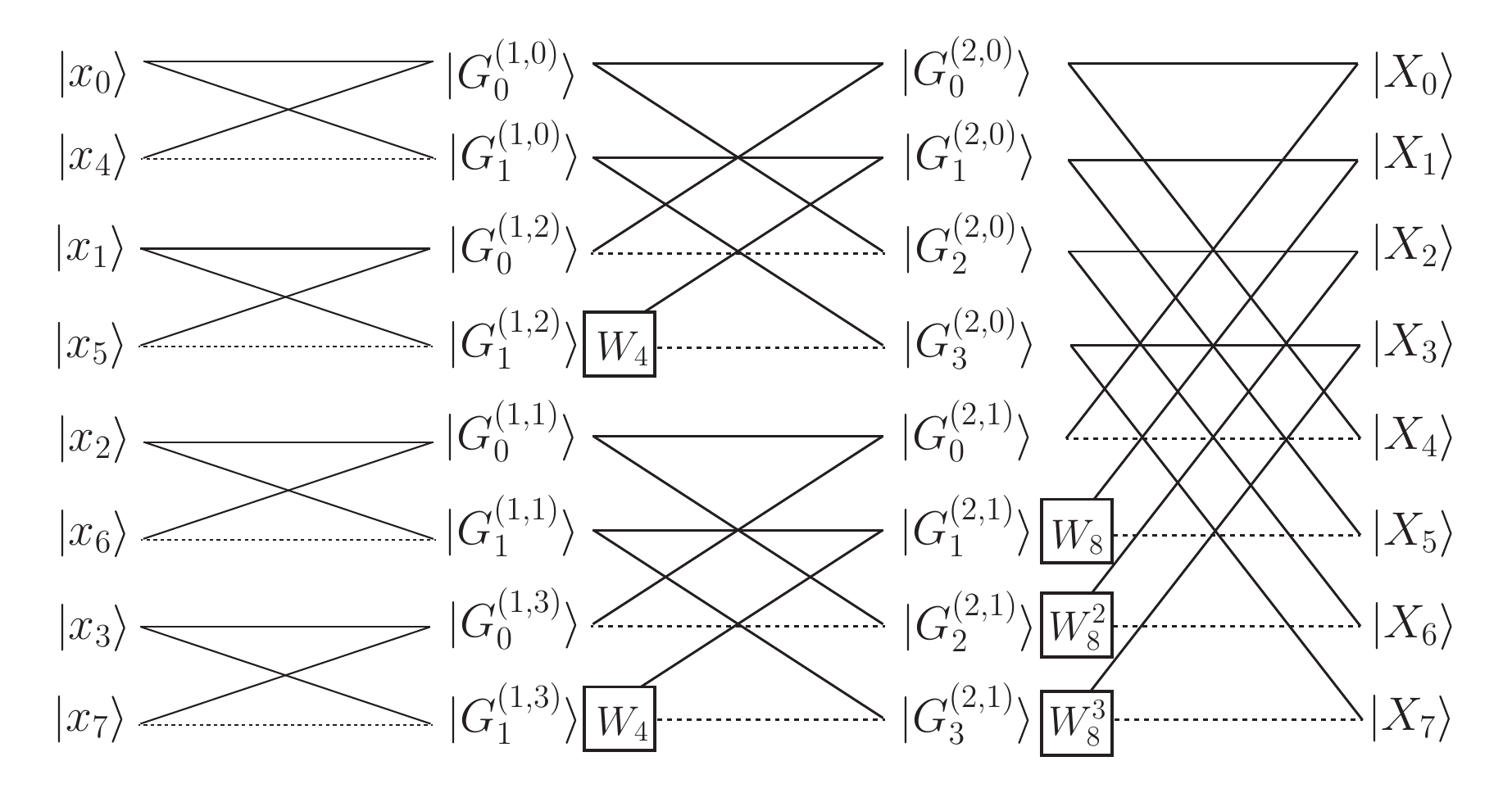}
  \caption{A pictorial representation of the QFFT
  $\bigotimes_{j=0}^{N-1}\ket{x_j}\longmapsto\bigotimes_{k=0}^{N-1}\ket{X_k}$ for $N=8$.
  Here, $W_{k}=\exp(-2\pi i/k)$, and
  $G^{(j,p)}_k$ is determined by the recursion relation \eqref{recursion} with
\eqref{initial}.
  }
  \label{8-FFT}
\end{figure}

Most importantly, 
the QFFT/FFT is decomposed into $\log_2N$ ``layers", 
where each layer consist of $N/2$ butterfly diagrams 
(see Fig.~\ref{8-FFT} for $N=8$): Totally 
$(N\log_2 N)/2$ diagrams are used in the QFFT/FFT.
As a result, the total computational complexity of the 
Fourier transform  \eqref{FT} is reduced from $O(N^2)$ to
$O(N\log_2 N)$ by the above procedure.

\subsection{Elementary operations in the QFFT}
As seen in \eqref{matrix}, to implement the QFFT in a quantum circuit,
the multiplication of the matrices
\begin{align}
  \left[\begin{matrix}1&1\\1&-1\end{matrix} \right],
  \quad
  \left[\begin{matrix}1&0 \\ 0&W_N^k\end{matrix}\right]
\end{align}
should be carried out in terms of quantum computation.
The first one is separated into an adder, a subtractor and
shift operators
by the LDU decomposition
\begin{align}
  \left[\begin{matrix}1&1\\1&-1\end{matrix}\right]
  =
  \left[\begin{matrix}1&0\\1&-1\end{matrix}\right]
  \left[\begin{matrix}1&0\\0&2\end{matrix}\right]
  \left[\begin{matrix}1&1\\0&1\end{matrix}\right].
\end{align}
Utilizing  the matrix-like notation as in
\eqref{matrix}, the action of the first matrix defined in \eqref{matrix}
on states $\ket{a}$ and $\ket{b}$
can be graphically interpreted as
\begin{equation}
\includegraphics[width=0.6\textwidth]{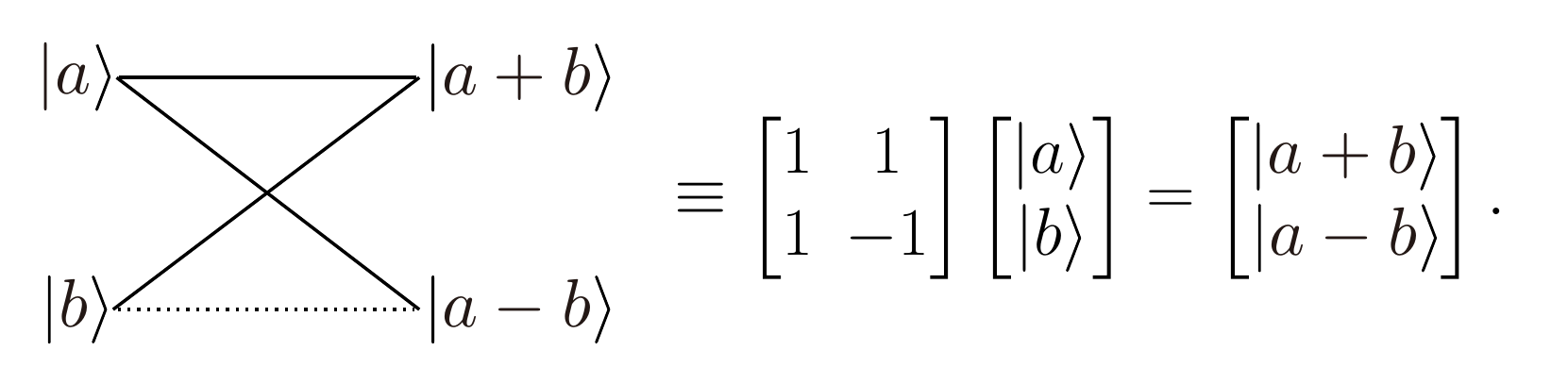}
\label{butterfly1}
\end{equation}
Note again that the above operation differs from the
conventional matrix operations.  On the other hand,
the second matrix is simply expressed as
\begin{equation}
\includegraphics[width=0.6\textwidth]{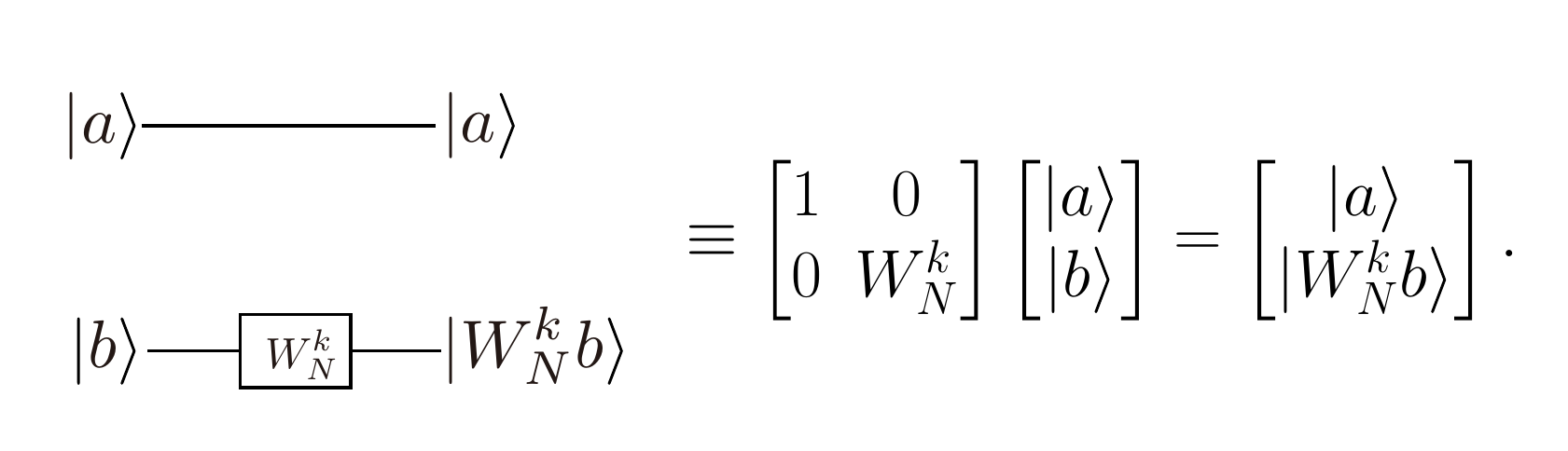}
\label{rot}
\end{equation}
Thus the butterfly diagram as in \eqref{FT-pic} or \eqref{matrix} can be written as
\begin{equation}
\includegraphics[width=0.8\textwidth]{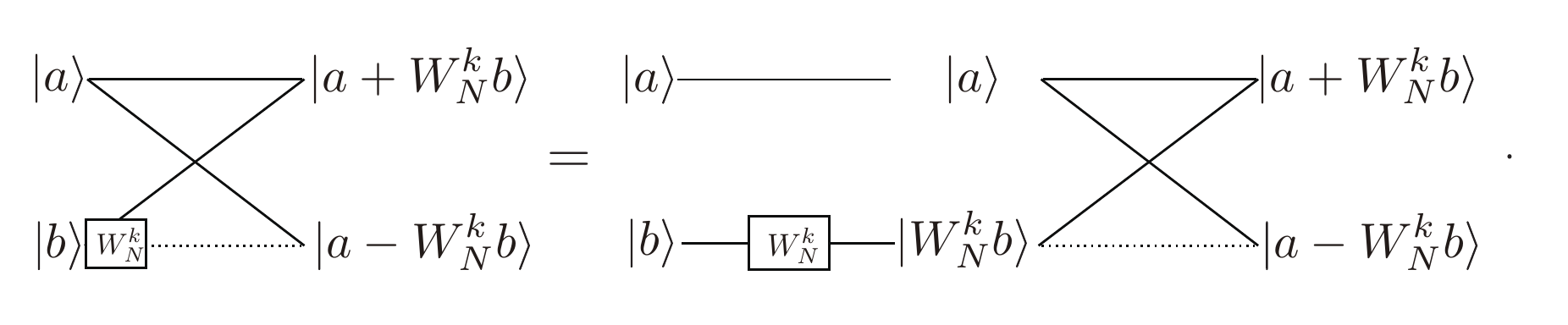}
\label{mult}
\end{equation}
Consequently, the QFFT can be implemented into a quantum circuit
consisting of adders, subtractors and shift operators.
In the next section, we explain these arithmetic
operators. An actual implementation of these operators into the butterfly
operations \eqref{mult} is deferred to Sect. 4.

\section{Quantum circuits for arithmetic operations}
\label{section:Quantum Ciruits for elementary caluculations}
%
%
In this section, we pictorially present a concept of some
quantum arithmetic operations such as a quantum adder, subtractor and
shift operators,  which are required to implement the QFFT as a quantum circuit.

Here, we adopt two's complement notation to represent a negative number.
Let us write a state $\ket{a}$ $(a\ge 0)$ using the binary representation
$\ket{a}=\ket{a_{n-1} \cdots a_0}:=\ket{a_{n-1}}\otimes \cdots \otimes \ket{a_0}$
($a_j\in\{0,1\}$). Let $m$ ($m>n$) be a total number of qubits of the system.
Let us  express $\ket{a}$ as
\begin{equation}
\ket{a}=\ket{\underbrace{a_{+}a_{+}\cdots a_{+}}_{m-n}a_{n-1}\cdots a_0},
\end{equation}
where $a_+=0$. Then, a negative number $\ket{b}$ ($=\ket{-a-1}$) can be
represented by the complement of $\ket{a}$:
\begin{equation}
\ket{b}=\ket{\underbrace{a_{-} a_- \cdots a_-}_{m-n}\bar{a}_{n-1}\cdots \bar{a}_{0}},
\end{equation}
where $a_-=\bar{a}_+=1$, $\bar{0}=1$ and $\bar{1}=0$. Namely, for the
$m$-qubit system,
the number $\{-2^{m-1},-2^{m-1}+1,\dots,2^{m-1}-1\}$ can be
expressed by this notation. For instance, $m=3$
\begin{alignat}{5}
&\ket{0}=\ket{000},& &\ket{1}=\ket{001},&&\ket{2}=\ket{010},&
&\ket{3}=\ket{011},\nn \\
&\ket{-4}=\ket{100},&  \quad &\ket{-3}=\ket{101},&\quad
&\ket{-2}=\ket{110},&\quad &\ket{-1}=\ket{111}.
\end{alignat}

\subsection{Sign extension}
\label{subsection:Extension of digit width}

In the actual computation, to avoid overflow, we sometimes
need to increase the number of bits (a so-called sign extension). This operation can be
achieved by just inserting $a_{\pm}$'s to the representation:
For instance, the representation for the $m$-qubit system
can be extended to that for the $l$-qubit system ($l>m$):
\begin{equation}
\ket{\underbrace{a_{\pm}\cdots a_{\pm}}_{m-n}a_{n-1}\cdots a_0}
\longmapsto
\ket{\underbrace{a_{\pm}\cdots a_{\pm}}_{l-n}a_{n-1}\cdots a_0}.
\end{equation}
In Fig.~\ref{figure:expansion}, we show a quantum circuit
to increase the number of digits from 4-qubit to 6-qubit.

In  Appendix, the number of extra qubits $a_{\pm}$ required 
for the QFFT is discussed.

\begin{figure}[t]
\centering
\includegraphics[width=0.35\textwidth]{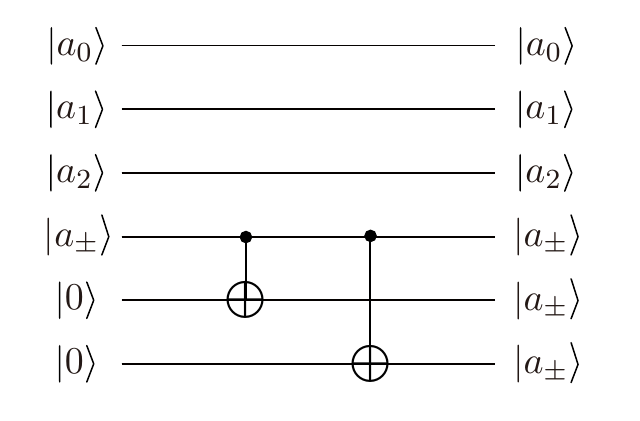}
\caption{An operation to increase the number of digits from $4$-qubit
$\ket{a_{\pm} a_2 a_1 a_0}$ to $6$-qubit
$\ket{a_{\pm} a_{\pm} a_{\pm} a_2 a_1 a_0}$,
where $a_+=0$ (resp. $a_-=1$) for a positive (resp. negative) number.
The extended state is achieved by copying $a_{\pm}$ via CNOT gates.}
\label{figure:expansion}
\end{figure}

\subsection{Adding and subtracting operations}
\label{adder}

\begin{figure}[ttt]
  \centering
  \includegraphics[width=0.4\textwidth]{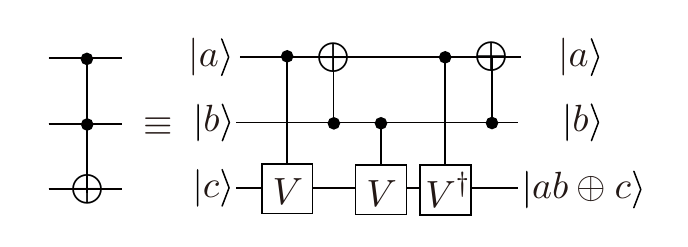}
\hspace*{1cm}
  \includegraphics[width=0.4\textwidth]{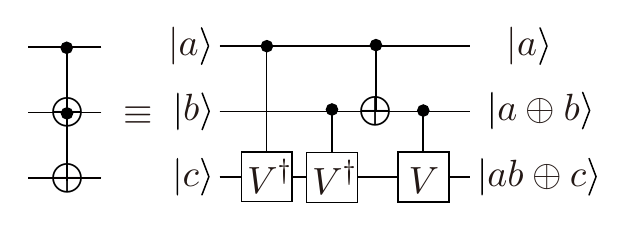}
  \caption{The Toffoli gate (left panel) and the Peres gate (right panel). 
$V$ and $V^{\dagger}$ are  defined by \eqref{V-mat}. The Toffoli and Peres
gates require 5 and 4 quantum gates, respectively.
}
  \label{fig-element}
\end{figure}

Let us consider an adder and a subtractor, by slightly modifying the
arguments developed in \cite{Q-Adder6, Q-Subb4}.
\begin{figure}[t]
  \centering
  \includegraphics[width=0.9\textwidth]{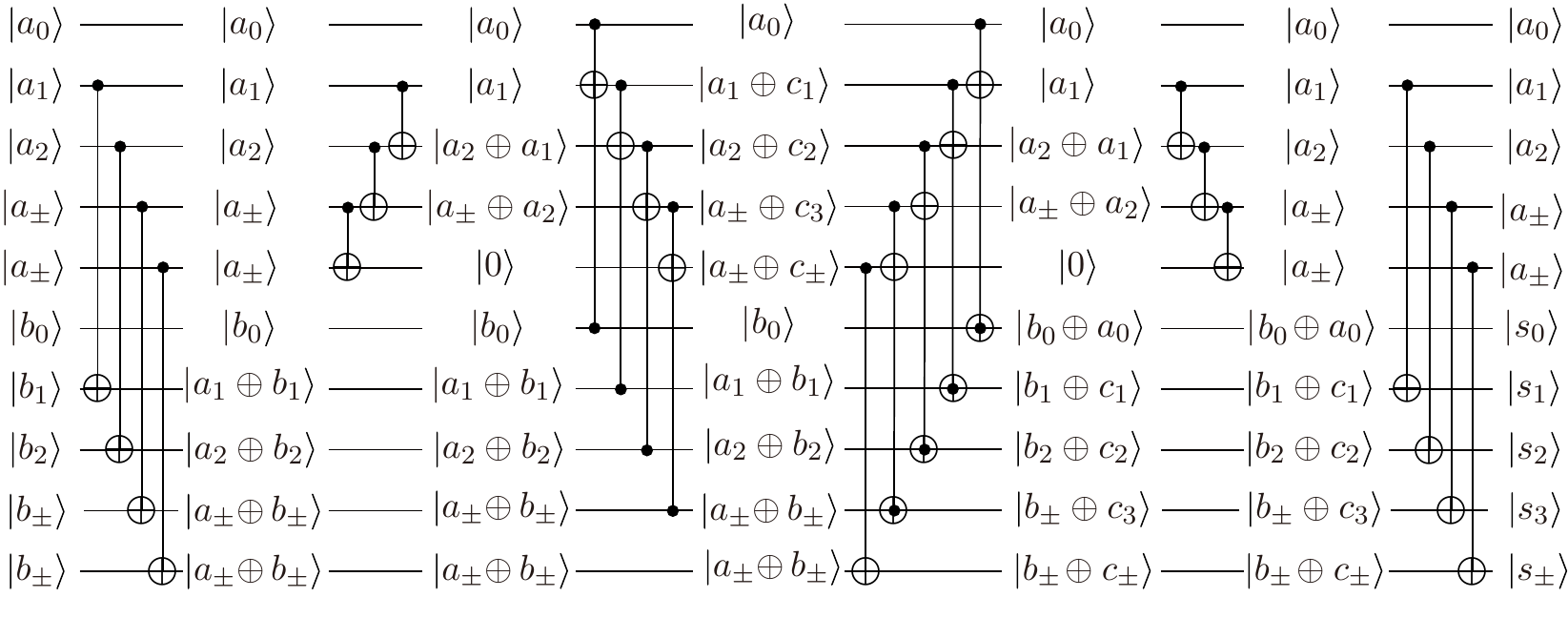}
  \caption{An adder circuit   for
  $\ket{a_{\pm} a_{\pm} a_2 a_1 a_0 + b_{\pm} b_{\pm} b_2 b_1 b_0}$. 
The circuit consists of the Toffoli and the Peres gates
defined as in Fig.~\ref{fig-element}. 
$c_j$ and $s_j$ ($j=1, 2, 3$) are the carry bit and the sum bit defined in \eqref{cs}.}
  \label{figure:imp_addCircuit_new}
\end{figure}
The addition of two $n$-bit numbers with the binary representation
$a=a_{n-1} \cdots a_0$ and $b=b_{n-1} \cdots b_0$
($a_j, b_j\in\{0,1\}$)
is calculated as 
\begin{equation}
 \includegraphics[width=0.5\textwidth]{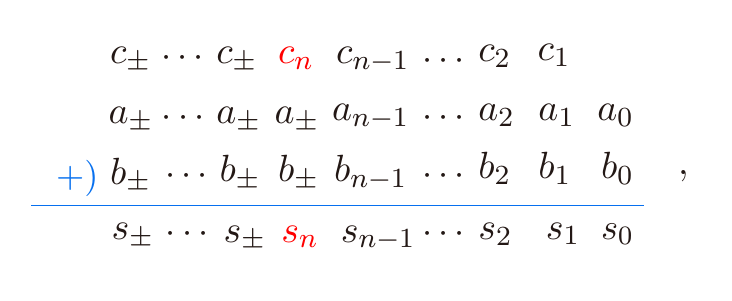}
\end{equation}
where the carry bit $c_j$ and the sum bit $s_j$ ($j=1, \cdots n$) are
defined by
\begin{align}
  c_j &=
  \begin{cases}
  a_0 b_0 &(j=1)
  \\
  a_{j-1}b_{j-1} \oplus b_{j-1}c_{j-1} \oplus c_{j-1}a_{j-1} &(2 \leq j \leq n)
  \end{cases},  \nn \\
  s_j &=
  \begin{cases}
  a_0\oplus b_0 &(j=0)\\
  a_j \oplus b_j \oplus c_j &(1 \leq j \leq  n-1)
  \\
  a_{\pm}\oplus b_{\pm}\oplus c_n &(j=n) 
  \end{cases}. \label{cs}
\end{align}
Note that the symbol $\oplus$ denotes exclusive disjunction.
In terms of a quantum circuit, this addition is implemented
in the transformation of the state
\begin{align}
  \ket{a} \otimes \ket{b}
  \longmapsto
  \ket{a} \otimes \ket{a+b},
\end{align}
and
graphically, it reads
\begin{equation}
\includegraphics[width=0.5\textwidth]{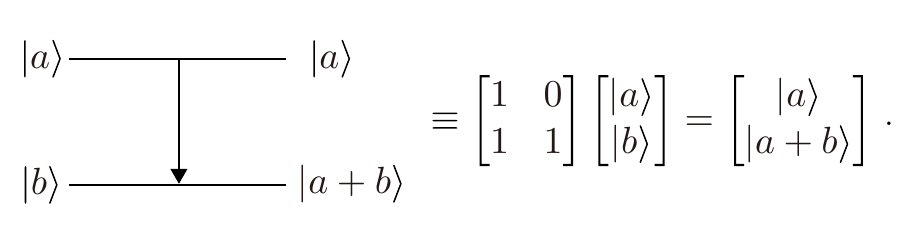}
\label{q-adder}
\end{equation}
Figure \ref{figure:imp_addCircuit_new} shows the actual circuit which is
a slightly modified version of a quantum adder originally developed in \cite{Q-Adder6}.
The adder 
circuit consists of the Toffoli gate \cite{Toffoli} and the Peres gate \cite{Peres}
defined as in Fig.~\ref{fig-element}, where $V$ and $V^{\dagger}$
are, respectively, given by
\begin{equation}
V=\frac{1+i}{2}\begin{pmatrix}1 & -i \\ -i & 1\end{pmatrix},
\quad
V^{\dagger}=\frac{1-i}{2}
\begin{pmatrix}1 & i \\ i & 1\end{pmatrix}.
\label{V-mat}
\end{equation}

On the other hand,  using the identity $\overline{\overline{a}+b} = a-b$,
we define a quantum subtractor as
\begin{align}
  \ket{a} \otimes \ket{b} \longmapsto
  \begin{cases}
    \ket{a} \otimes \ket{\overline{ \overline{a}+b }}
    = \ket{a} \otimes \ket{a-b}
    \\
    \ket{a} \otimes \ket{\overline{ a+\overline{b} }}
    = \ket{a} \otimes \ket{-a+b}
  \end{cases},
\end{align}
which can be implemented by just inserting NOT gates (denoting it by the symbol 
$\bigoplus$) to the
above defined adder \eqref{q-adder} \cite{Q-Subb4}: 
\begin{align}
&   \includegraphics[width=0.8\textwidth]{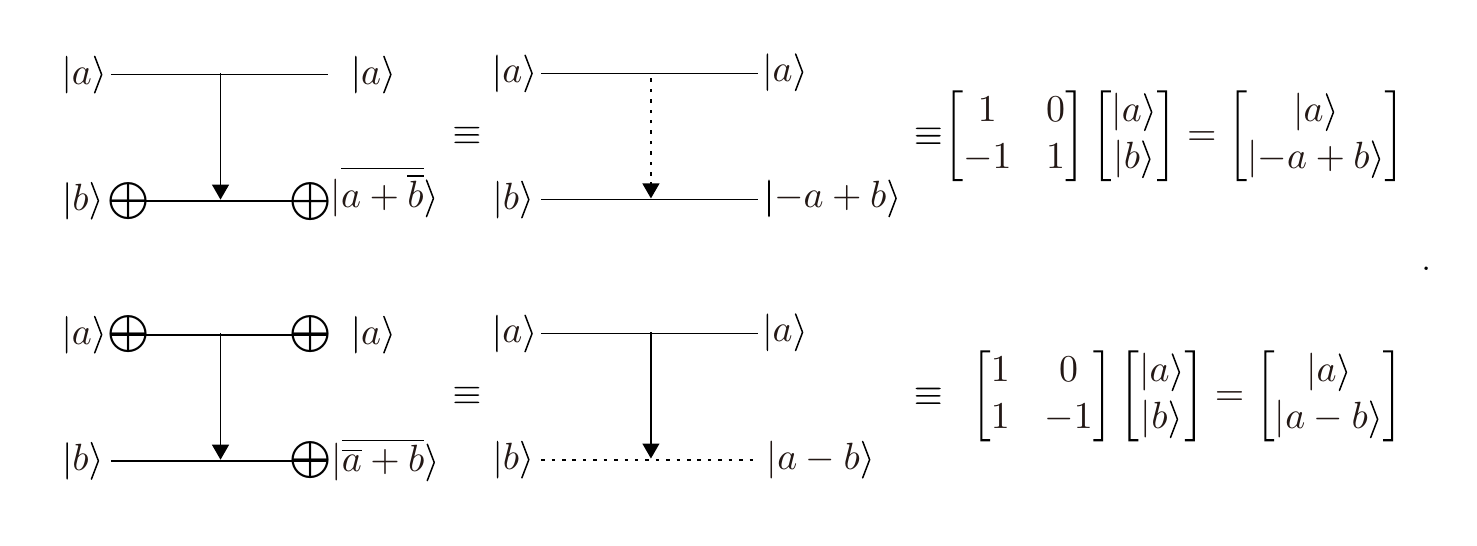} 
\label{q-subtractor}
\end{align}

The quantum circuit of the adder for $n_{\rm in}$-qubit input data consists
of 6 ``layers" as in Fig.~\ref{figure:imp_addCircuit_new}.
(Note here that the number of the layers does
not depend on $n_{\rm in}$.) The first, second, fifth and sixth layers,
respectively, contain $n_{\rm in}-1$, $n_{\rm in}-2$,
$n_{\rm in}-2$ and $n_{\rm in}-1$ CNOT gates. The third layer
consists of $n_{\rm in}-1$ Toffoli gates: $5(n_{\rm in}-1)$ CNOT gates
are required. The fourth layer contains $n_{\rm in}-1$ Peres gates and
one CNOT gates: $4 (n_{\rm in}-1)+1$ CNOT gates are required. Note
that the Toffoli (resp. Peres) gate requires 5 (resp. 4) CNOT gates
as shown in Fig.~\ref{fig-element}.
Thus, totally $13 n_{\rm in}-14$ quantum gates are required for the adder circuit
for $n_{\rm in}$-qubit data. On the other hand, the subtractor 
defined by \eqref{q-subtractor} requires additional at most 
$3 n_{\rm in}$ CNOT gates, and hence,  totally at most $16 n_{\rm in}-14$
quantum gates are required for the subtractor.

\subsection{Sign changing operation}

Due to the identity
\begin{align}
  -a=(-1)\times a=\overline{a}+1,
\end{align}
we can change the sign of the input number
by an adder with NOT gate:
\begin{equation}
\includegraphics[width=0.45\textwidth]{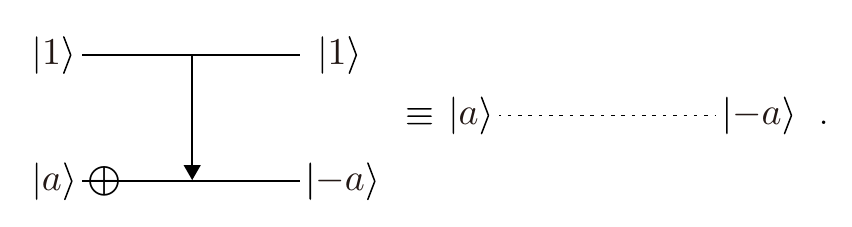}
\end{equation}
\subsection{Arithmetic shift operations}
Let us implement an operation to multiply by $2^p$ ($p\in \mathbb{N}$):
\begin{equation}
\ket{a} \longmapsto \ket{2^p a}.
\end{equation}
This operation is carried out by shifting the digits to
the left (arithmetic left shift):
\begin{equation}
\ket{a}=\ket{\underbrace{a_{\pm}\dots a_{\pm}}_{m-n} a_{n-1}\dots a_0}\longmapsto
\ket{\underbrace{a_{\pm}\dots a_{\pm}}_{m-n-p} a_{n-1}\dots a_0 \underbrace{0\cdots 0}_p}
=\ket{2^p a}.
\end{equation}
Let us pictorially express this operation as
\begin{equation}
\includegraphics[width=0.5\textwidth]{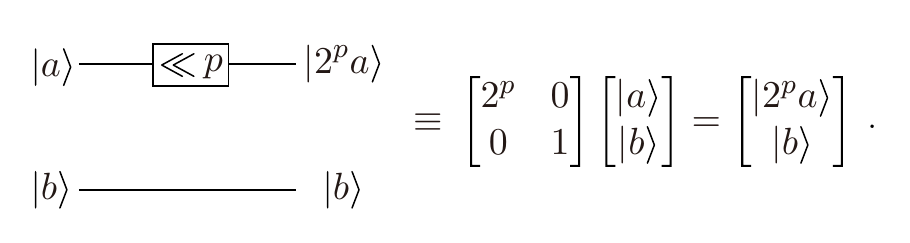}
\end{equation}

In a similar manner, we can define an arithmetic right shift  which is an operation
to multiply by $2^{-p}$:
\begin{align}
\ket{a}=&\ket{\underbrace{a_{\pm}\dots a_{\pm}}_{m-n} a_{n-j-1}
\dots a_{0}a_{-1}\cdots a_{-j}}\nn \\
&\longmapsto
\ket{\underbrace{a_{\pm}\dots a_{\pm}}_{m-n+p} a_{n-j-1}\cdots a_0a_{-1}\cdots a_{-j+p}}
=\ket{2^{-p} a},
\label{right-shift1}
\end{align}
where $a_{-k}$ $(1\le k \le j-p)$ are the fractional part of $a$. Note that, in the
above operation,
$p$ significant  digits are lost. We also graphically denote this operation
\begin{equation}
\includegraphics[width=0.5\textwidth]{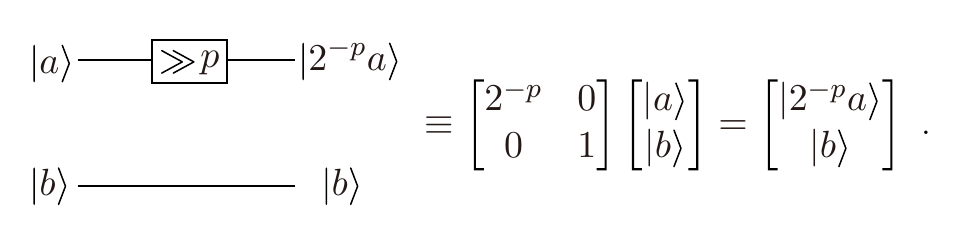}
\label{right-shift2}
\end{equation}

\begin{figure}[t]
\centering
\includegraphics[width=0.8\textwidth]{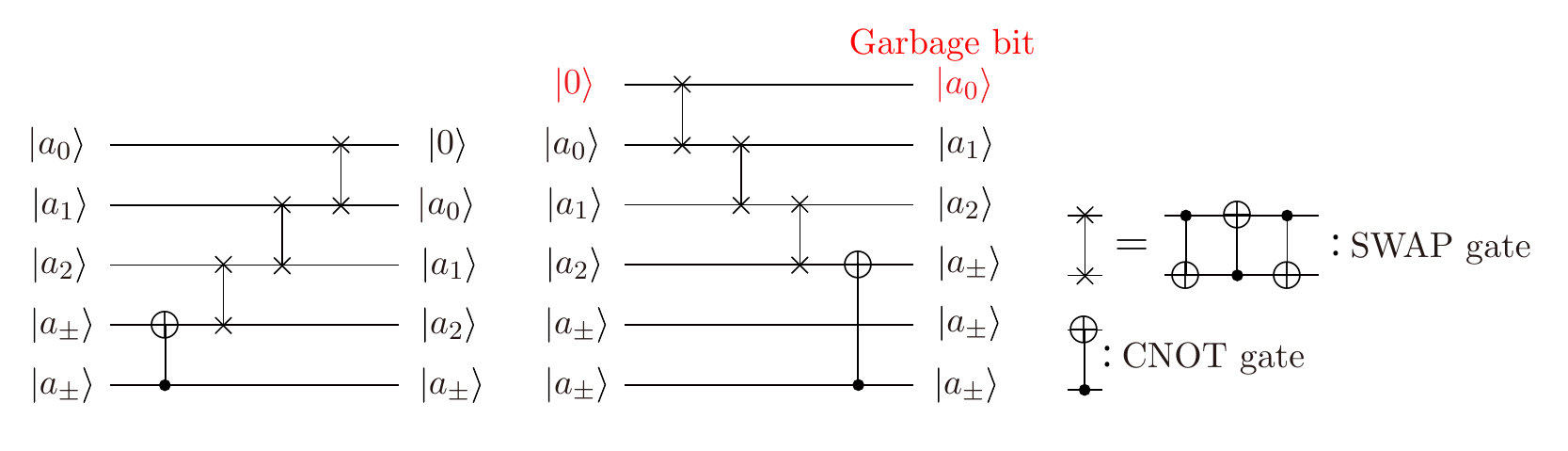}
\caption{
The left (right) panel denotes a quantum circuit for an arithmetic left 
shift (right shift). 
The least significant digit $a_0$ is lost for this  right shift 
operation (right panel). Totally  $3n_{\rm in}-5$ quantum 
gates (one CNOT gate and $n_{\rm in}-2$ SWAP gates containing 3 CNOT 
gates) are required 
for the shift operation of  $n_{\rm in}$-qubit input data.}
\label{shift-op}
\end{figure}

The actual implementation of these shift operations into quantum circuits
can be accomplished by certain combinations of SWAP and CNOT gates:
 $3n_{\rm in}-5$ quantum 
gates (one CNOT gate and $n_{\rm in}-2$ SWAP gates consisting of 3 CNOT gates) 
are required for the shift operation of $n_{\rm in}$-qubit input data.
In Fig.~\ref{shift-op},
we show a quantum circuit for the left (right) shift operation
for  $m=3$, $n=3$ and $p=1$.

Combining  the adder, the subtractor and the shift operations,
we can arithmetically manipulate an arbitrary number.

\section{Decomposition of the butterfly operation}
\label{section:Quantum Circuits for FFT}
\begin{figure}[t]
  \centering
  \includegraphics[width=0.7\textwidth]{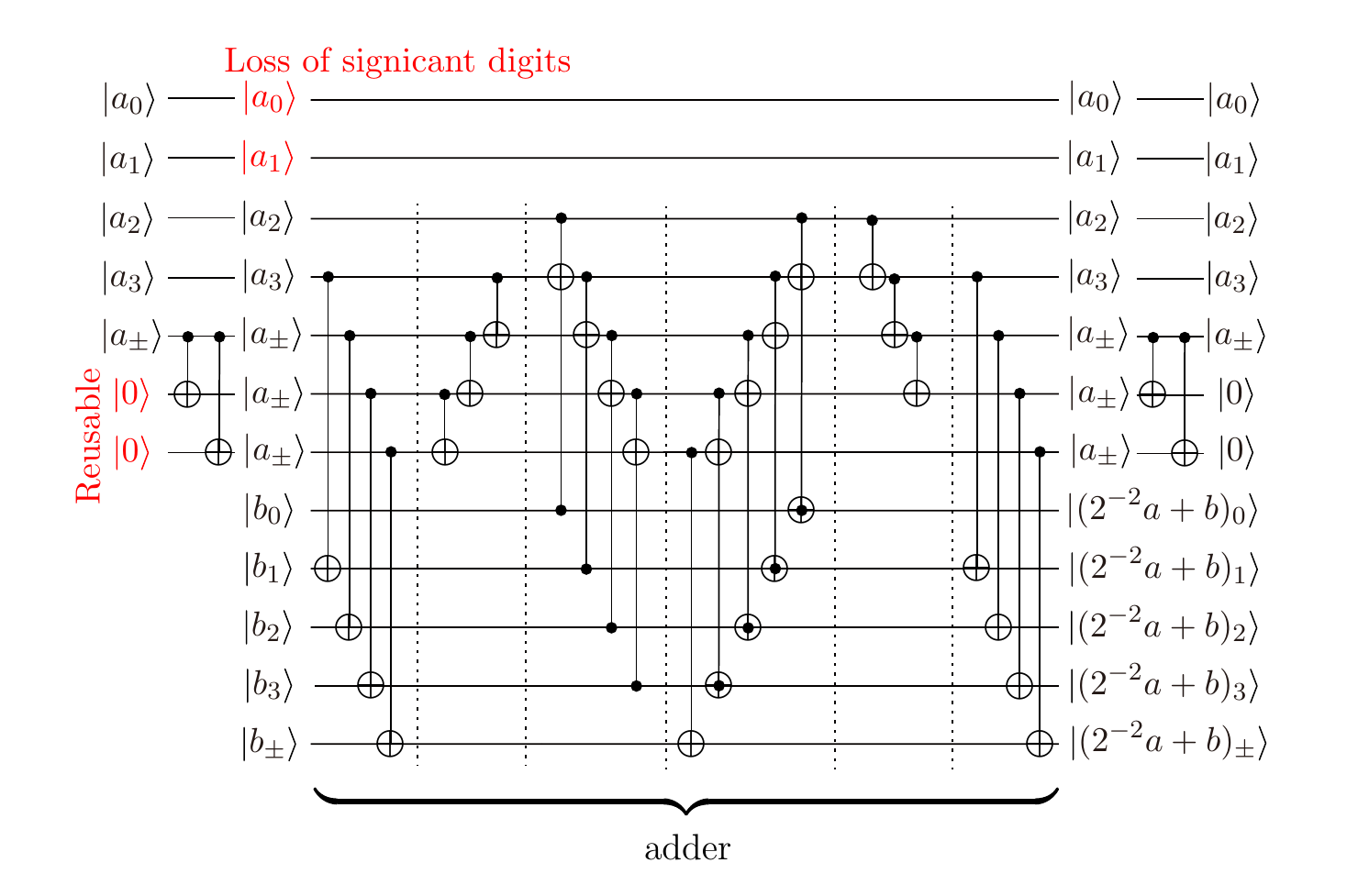}
  \caption{A quantum circuit of an adder with a right shift operation
\eqref{add-shift} for $p=2$. Though two additional qubits are required
for this right shift operation, these two qubits are reusable. Moreover 
this quantum circuit does not generate any garbage bits. 
Except for this shift operation, the circuit essentially consists of 
the adder requiring $13 n_{\rm in}-14$ quantum gates
for  $n_{\rm in}$-qubit input data.
Some elements
of the adder circuit is defined in Fig.~\ref{fig-element}.}
  \label{fig:shiftcalc_def}
\end{figure}

Now, we decompose the butterfly operation \eqref{mult} (see also
\eqref{FT-pic}), which plays a
central role on the QFFT, into the elementary arithmetic
operations shown in previous section.
First, we  decompose the butterfly operation into elementary
operations: 
\begin{equation}
  \includegraphics[width=0.6\textwidth]{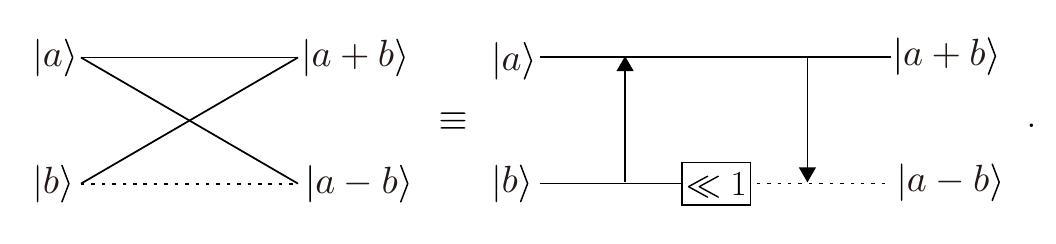}
\label{butterfly}
\end{equation}
The above operation \eqref{butterfly} is implemented as a quantum circuit
consisting of one adder,  one subtractor (see \eqref{q-adder}, \eqref{q-subtractor}
and Fig.~\ref{figure:imp_addCircuit_new} in Sect.~\ref{adder} 
for a quantum circuit for the adder/subtractor) and one 
shift operation \eqref{right-shift2} (see also Fig.~\ref{shift-op}),
which, respectively, require $13 n_{\rm in}-14$, $16n_{\rm in}-14$ and
$3n_{\rm in}-5$ quantum gates for $n_{\rm in}$-qubit input.
Therefore, the number of quantum gates used in the implementation of \eqref{butterfly}
is totally $32 n_{\rm in}-33$.

In the butterfly operation, the input states consist of $\ket{(W_N^k) a}$.
Let us abbreviate the component $W_N^k = \exp\left(
-i \frac{2\pi}{N}k \right)$
to $\exp \left(i\theta \right)$ for simplicity.
The calculation of $\ket{\exp \left(i\theta \right)a}$ is decomposed into
\begin{align}
   &\ket{\exp(i\theta)a} = \ket{(\cos\theta+i\sin\theta)(a_r+ia_i)}
   = \begin{bmatrix}
        1&i
     \end{bmatrix}
     \begin{bmatrix}
       \cos\theta&-\sin\theta\\\sin\theta&\cos\theta
     \end{bmatrix}
     \begin{bmatrix}
         \ket{a_r}\\\ket{a_i}
     \end{bmatrix},
\end{align}
where $a_r$ and $a_i$ are, respectively, the real and imaginary part of $a$.
The rotation matrix is further decomposed into adding (resp. subtracting) 
operators \eqref{q-adder} (resp. \eqref{q-subtractor})  
with arithmetic right shift operations \eqref{right-shift2} \cite{FFT-Int,FFT-app,
Wavelet}:
\begin{empheq}[left={
\begin{bmatrix}
       \cos\theta&-\sin\theta\\\sin\theta&\cos\theta
     \end{bmatrix}=\empheqlbrace}]{alignat=2}
&\begin{bmatrix}1&\frac{\cos\theta-1}{\sin\theta}\\0&1\end{bmatrix}
            \begin{bmatrix}1&0\\\sin\theta&1\end{bmatrix}
              \begin{bmatrix}1&\frac{\cos\theta-1}{\sin\theta}\\0&1\end{bmatrix} \label{up}\\
         &\begin{bmatrix}-1&\frac{\cos\theta+1}{\sin\theta}\\0&1\end{bmatrix}
            \begin{bmatrix}1&0\\\sin\theta&-1\end{bmatrix}
              \begin{bmatrix}1&\frac{\cos\theta+1}{\sin\theta}\\0&1\end{bmatrix}\label{down}.
\end{empheq}
Because
\begin{equation}
  \left|\frac{\cos\theta-1}{\sin\theta} \right|\le 1 \,\,
\text{for $\theta\in[-\frac{\pi}2,\frac{\pi}2]$},\quad
\left|\frac{\cos\theta+1}{\sin\theta} \right|< 1 \,\,
\text{for $\theta\in[-\pi,-\frac{\pi}2)\cup(\frac{\pi}2,\pi]$},
\end{equation}
to apply the right shift operator,
we use \eqref{up} for $\theta\in[-\frac{\pi}2,\frac{\pi}2]$, while  for
$\theta\in[-\pi,-\frac{\pi}2)\cup(\frac{\pi}2,\pi]$, we use \eqref{down}.
The each matrix in the RHS of the above decomposition is schematically
given by
\begin{equation}
  \includegraphics[width=0.3\textwidth]{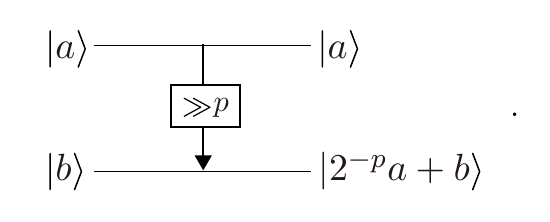}
\label{add-shift}
\end{equation}
In fact, this decomposition makes it possible to
efficiently implement the elementary arithmetic operations
required for the QFFT. Namely, we develop an implementation of
the above procedure into a quantum circuit so as not
to generate any garbage bits. 
A quantum circuit for the operation \eqref{add-shift} can be
constructed by the adder circuit given in Sect.~\ref{adder} 
(see Fig.~\ref{figure:imp_addCircuit_new}).
For $p=2$, the circuit is  shown in Fig.~\ref{fig:shiftcalc_def}. 
Except for some sign extension, the implementation of this
operation requires $13 n_{\rm in}-14$  quantum gates,
which is the number of the quantum gates in
the adder  for  $n_{\rm in}$-qubit input data 
(see Sect.~\ref{adder} in detail).

Thanks to the circuit \eqref{add-shift}, we can construct 
quantum circuits of adding and subtracting for an arbitrary number:
\begin{align}
&\includegraphics[width=0.6\textwidth]{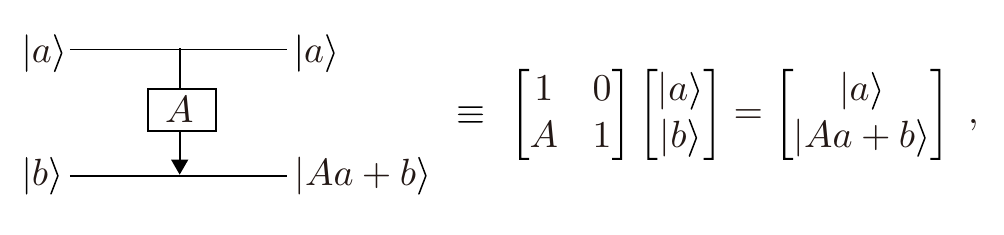} \nn \\
&\includegraphics[width=0.6\textwidth]{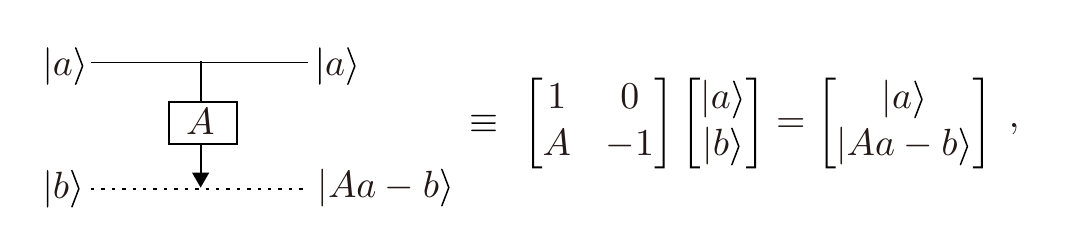} \\
&\includegraphics[width=0.6\textwidth]{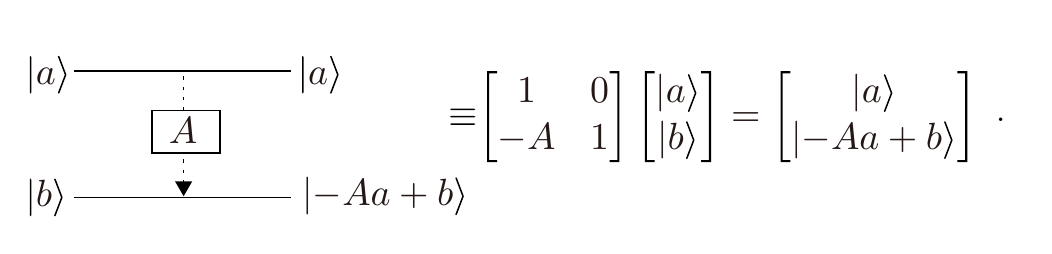} \nn 
\end{align}
For instance, to add $A=7\times a$ to $b$, we just apply 
the add operation \eqref{add-shift} three times, since
\begin{equation}
7\times a=a\ll 2+a\ll 1+a.
\end{equation}
Thus, the decompositions \eqref{up} and \eqref{down} are
summarized  graphically: 
\begin{equation}
  \includegraphics[width=0.7\textwidth]{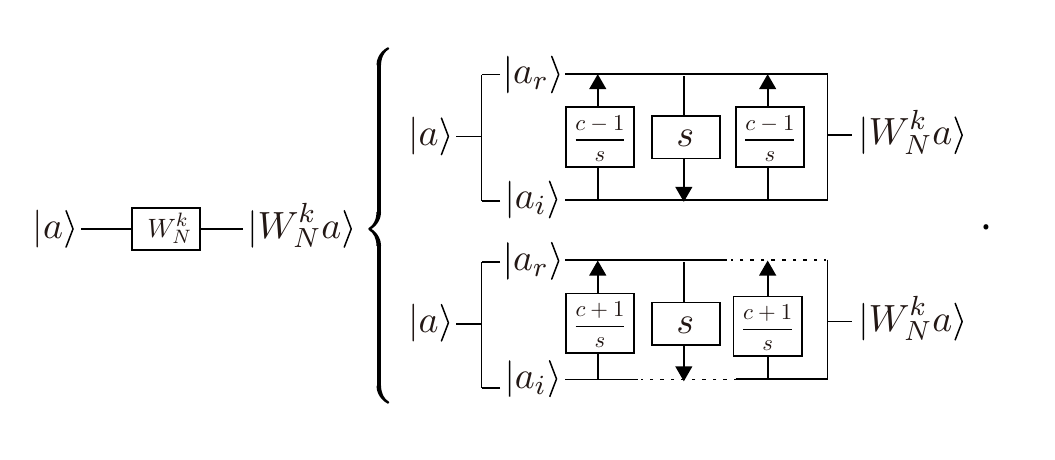}
\label{rotation}
\end{equation}
Note that the quantum circuit of the first (resp. second) operation
in \eqref{rotation} consists of three adders (resp. one adder and 
two subtractors). If we require an accuracy of $2^{-A}$ for the rotation
$|W_N^k|$, then the quantum circuit needs $A\times 3\times (13n_{\rm in}-14)$
(resp. at most $A \left\{ 2(16n_{\rm in}-14)+13n_{\rm in}-14\right\}$
for the first (resp. second) operation of $n_{\rm in}$-qubit input.

In summary, the butterfly operation \eqref{FT-pic}, which plays
a central role for the QFFT (Fig.~\ref{8-FFT}), is decomposed
into two operations  in \eqref{mult}. The first operation in
\eqref{mult} is further divided  as shown in \eqref{butterfly}
which requires  $32 n_{\rm in}-33$ quantum gates for  $n_{\rm in}$-qubit 
input. The second operation in \eqref{mult} can be
reduced to  \eqref{rotation} consisting of at most $A(45n_{\rm in}-42)$
quantum gates. As a consequence, the butterfly operation \eqref{FT-pic}
consists of at most $32 n_{\rm in }-33+A(45 n_{\rm in }-42)$ quantum gates. 
The number of quantum gates of the main operations necessary for 
the QFFT is summarized in Table~\ref{table}.

\begin{table}[ttt]
\centering
  \begin{tabular}{|l|l|} \hline
  Operation & Number of quantum gates  \\ \hline 
    Adder \eqref{q-adder} (Fig.~\ref{figure:imp_addCircuit_new})  &  $13n_{\rm in}-14$ \\
    Subtractor \eqref{q-subtractor} &  $16n_{\rm in}-14$ \\
    Shift \eqref{right-shift2} (Fig.~\ref{shift-op}) & $3n_{\rm in}-5$  \\ 
    Butterfly \eqref{butterfly1} &    $32n_{\rm in}-33$\\ 
    Rotation \eqref{rot}/\eqref{rotation} & $A(45n_{\rm in}-42)$ \\
    Butterfly \eqref{mult} &   $32n_{\rm in}-33+A( 45n_{\rm in}-42)$ \\ \hline
  \end{tabular}
\caption{The (maximum) number of quantum gates required for some elementary 
operations for $n_{\rm in}$-qubit input. $2^{-A}$ denotes the computational
accuracy of the rotation $|W_N^k|$ in \eqref{rotation}.}
\label{table}
\end{table}

Considering that quantum circuit is reversible, the calculation of inverse QFFT 
requires two matrices
\begin{align}
  \left\{ \left[\begin{matrix}1&1\\1&-1\end{matrix}\right] \left[\begin{matrix}1&0\\0&(W_N)^k\end{matrix}\right] \right\}^{-1}
  =\left[\begin{matrix}1&0\\0&(W_N)^{-k}\end{matrix}\right]  \left[\begin{matrix}1/2&1/2\\1/2&-1/2\end{matrix}\right],
\end{align}
which are also decomposed into an adder, a subtractor and  shift operators.

\section{Computational complexities}
\label{cost}
\subsection{Total number of quantum gates}
The QFFT algorithm described in Sect. 2 is decomposed 
into several arithmetic operations, which is implemented 
in quantum circuits as  in Sect. 3 and 4.
Here, we  estimate the total number of the quantum gates required 
for the implementation. As  in Sect. 2, the QFFT consists
of $(N\log_2 N)/2$ butterfly operations, where each operation 
consists of at most $32 n_{\rm in }-33+A(45 n_{\rm in }-42)$ quantum gates
as explained in Sect. 4. Here, $n_{\rm in}$ and $A$ denote the
number of input qubits and the accuracy of rotation, respectively. See Table~\ref{table}
for the number of quantum of some elementary operations.
As a result, the total number
of the gates $n_{g}$ required for the QFFT is estimated to be at most
\begin{equation}
n_g=\left\{32 n_{\rm in }-33+A(45 n_{\rm in }-42)\right\}\times \frac{N}{2}\log_2 N.
\end{equation}
\subsection{Computational costs including data encoding}
To take advantage of quantum computing, some efficient method to encode classical
data in quantum states such as a qRAM \cite{qRAM1,qRAM2,qRAM3} must be necessary. 
Here, we briefly comment on  computational costs for the QFFT including 
data encoding, taking a simple example as illustrated in the introduction. 

Let us process $N$ images of  $L\times L$ pixels each 
(see Fig. \ref{gray} for $L=2$, for instance). For comparison, first, we analyze 
the computational complexity of the classical FFT. 
As described in the introduction, the  complexity to process
each image is $O(L^2 \log_2 L^2)$. Namely, the total cost
for the FFT is $O(N L^2 \log_2 L^2)$.

For the QFFT, one must 
encode the classical data stored in the classical 
RAM one by one in quantum states:  $O(N L^2)$ processes
are required to encode them. The computational
complexity of the QFFT to process the quantum images 
is $O(L^2 \log_2 L^2)$.
Consequently, the total complexity including data encoding
is $O(NL^2)+O(L^2\log_2 L^2)$. Thus, as long as we use the
classical RAM, there is not so much advantage.

Recently, a concept of quantum random access memory (qRAM)
which makes it possible to drastically reduce the computational 
cost to encode classical data has been
developed in \cite{qRAM1,qRAM2,qRAM3}: The complexity 
to encode the data can be reduced from $O(N L^2)$ 
to $O(L^2)$. Namely, the total computational complexity
for the QFFT with the qRAM is $O(L^2\log_2 L^2)$
which is much less than the conventional method.

In table~\ref{encoding-cost}, the complexities discussed here
are summarized.

\begin{table}[ttt]
\centering
  \begin{tabular}{|l|l|} \hline
   & Computational complexity  \\ \hline 
    FFT  &  $O(N L^2\log_2 L^2)$ \\
    QFFT+RAM &  $O(N L^2)+O(L^2\log_2 L^2)$ \\
    QFFT+qRAM& $O(L^2\log_2 L^2)$  \\ \hline
  \end{tabular}
\caption{Computational complexities to process 
 $N$ images of $L\times L$ pixels each by
the FFT, and the QFFT including  data encoding through the RAM and 
the qRAM.
}
\label{encoding-cost}
\end{table}

\section{Quantum information processing based on the QFFT}

As mentioned previously, one of the advantages of the QFFT
is its wide utility: The method is applicable to all the 
problems processed by the conventional FFT. Moreover, the QFFT can
simultaneously process  multiple data sets which can be 
generated by $U(N)$ transformations realized by quantum 
gates as in \cite{UN-gate}.

\begin{figure}[ttt]
  \centering
  \includegraphics[width=0.9\textwidth]{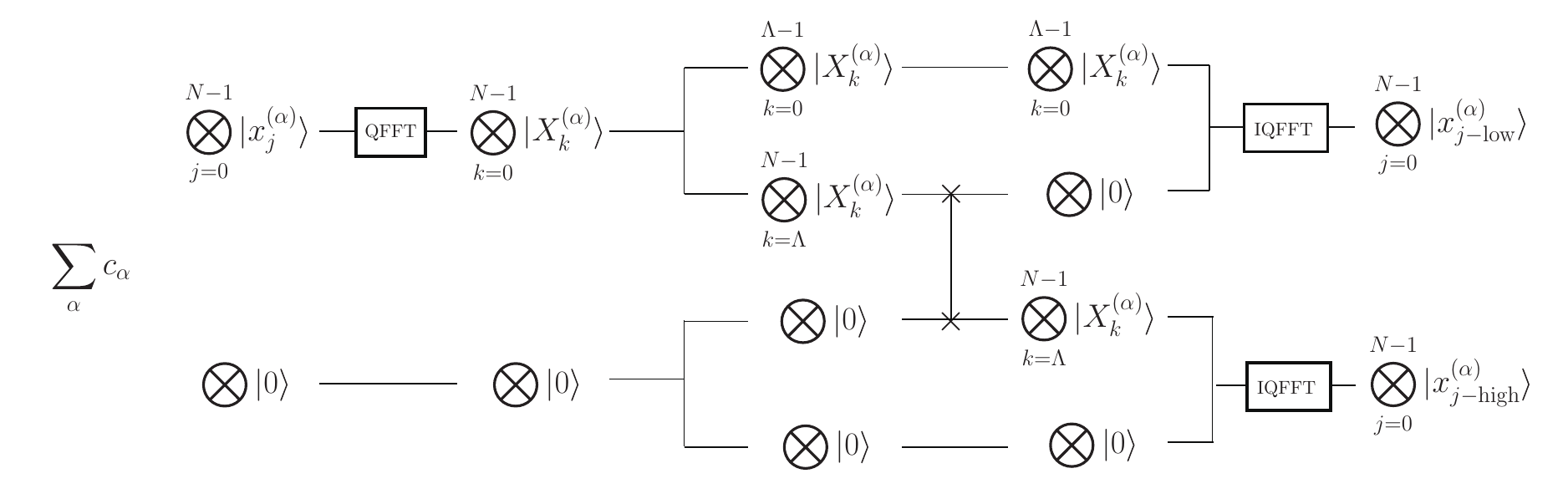}
  \caption{A quantum circuit for the high/low pass filter applying 
multiple data sets stored in a superposition state with amplitudes $\{c_\alpha\}$, where
  $\bigotimes_{j=0}^{N-1}\ket{x_j^{(\alpha)}}$ with an auxiliary state $\bigotimes \ket{0}$
   stands for a single $n$-qubit data ($N=2^n$) labeled $\alpha$.
  Swapping the high-frequency data (higher than a cutoff frequency $\Lambda$) 
  $\bigotimes_{k=\Lambda}^{N-1}\ket{X_k^{(\alpha)}}$
    with a part of the auxiliary state $\bigotimes \ket{0}$ and performing
the inverse QFFT (IQFFT),
we obtain both the high and low pass filtered data sets.}
  \label{fig:freq_div}
\end{figure}

\begin{figure}[t]
  \centering
  \includegraphics[width=0.85\textwidth]{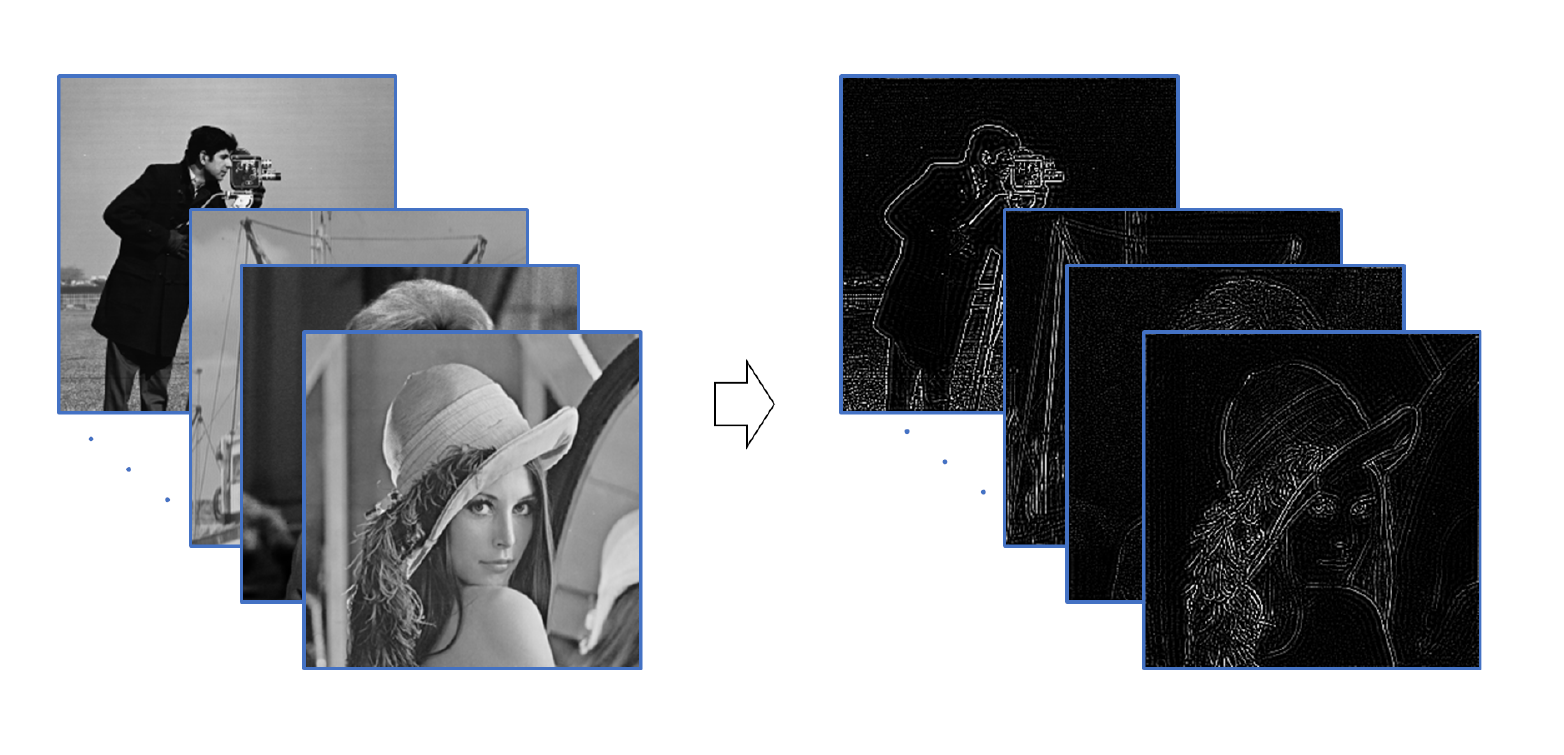}
  \caption{A conceptual image of the high pass filter applied to quantum 
multiple images.}
  \label{fig:hpf}
\end{figure}

As a concrete example, in Fig.~\ref{fig:freq_div}, we illustrate
a quantum circuit for the high/low pass filter
applying to  multiple data sets.
A single $n$-qubit data labeled $\alpha$ is described as 
$\bigotimes_{j=0}^{N-1}\ket{x_j^{(\alpha)}}$ ($N=2^n$) 
with an auxiliary state $\bigotimes \ket{0}$.
A sequence of operations, the QFFT, the SWAP gate acting on multiple qubits,
and the inverse QFFT (IQFFT), generates both the high and low pass filtered data sets 
separated with some cutoff frequency $\Lambda$ through a single circuit.
Multiple data sets can be processed simultaneously when the corresponding states are stored in a
superposition state with some probability amplitudes $\{c_\alpha\}$.
If enough numbers of data sets are given, this 
information processing system exceeds the one using the QFT.
Replacing the QFFT with the two-dimensional QFFT, we can also use this system as an edge detector
for multiple quantum images (see Fig.~\ref{fig:hpf} as
a conceptual image).

\section{Summary and Discussion}
In this paper, we have discussed an implementation of the
FFT as a quantum circuit.
The quantum version of the FFT (QFFT) is defined as
a transformation of a tensor product of quantum states.
The QFFT has been constructed by a combination of several fundamental
arithmetic operators such as an adder, subtractor and shift operators
which have been implemented into the quantum circuit of
the QFFT without generating any garbage bits.

One of the advantages of the QFFT is due to its high versatility:
The QFFT is applicable to all the problems that can be solved by
the conventional FFT.  For instance, the frequency domain 
filtering of digital images is one of the possible applications of the QFFT.
A major advantage of using the QFFT lies in its quantum superposition:
Multiple images are processed simultaneously.
It is even superior to the QFT when the number of
images is sufficiently large. 

Utilization of the resultant multiple data sets obtained after performing the QFFT is 
also interesting.
The QFFT sustains all the information of Fourier coefficients
until the moment the quantum state is measured.
If the quantum state that contains the Fourier coefficients of multiple data sets 
was passed on to some quantum device directly and there were some proper 
techniques to handle it, it would play a key role in the field of quantum 
machine learning.

\section*{Acknowledgment}

The present work was partially supported by Grants-in-Aid for Scientific
Research (C) Nos. 16K05468 and 20K03793 from the Japan Society for the 
Promotion of Science.

\begin{appendix}

%
\section{Number of extra qubits $a_\pm$ required for the QFFT}
\label{Appendix:Number of empty operand required for QFFT}
%
The key structure of the QFFT is the butterfly operation combined with
the multiplication of $W_N^k$.
In the whole QFFT process, we have $\log_2 N$ layers of this
structure.
By counting the ratio of the possible value ranges of 
the input and output of this structure, and multiplying it by $\log_2 N$,
we can estimate the number of qubits $a_\pm$ required for the QFFT.

Let $I_r$ and $I_i$ be the real and imaginary part of 
the input $I$ and impose the following 
conditions on $I$:
\begin{align}\label{align:I_condition}
  \begin{cases}
    -2^{n} \leq I_r,I_i \leq 2^n-1
    \\
    \qquad I_r^2 + I_i^2 \leq 2^{2n}
  \end{cases}.
\end{align}
Because the initial input is a real number ($I_i=0$), it satisfies
the above conditions.
The multiplication of $W_N^k$ is essentially a rotation
in the complex space.
Namely, it preserves the distance:
\begin{align}\label{align:WI_condition}
  [WI]_r^2 + [WI]_i^2 \leq 2^{2n},
\end{align}
where the suffixes $i$ and $r$ represent the real and imaginary part,
respectively.
The output $O$ of the butterfly structure then satisfies
\begin{align}\label{align:A_condition}
  \begin{cases}
    -2^{n+1} \leq O_r,O_i \leq 2^{n+1}-1
    \\
    \qquad O_r^2 + O_i^2 \leq 2^{2(n+1)}
  \end{cases}.
\end{align}
The absolute value of the output is doubled compared to that of the input.
Therefore, one extra qubit $a_\pm$ is required for one layer and
$\log_2 N$ $a_\pm$'s for the total QFFT.

\end{appendix}

\end{document}